\documentclass[twocolumn]{aastex631}
\usepackage{CJK}
\usepackage{threeparttable}
\usepackage{array}
\usepackage{multirow}
\usepackage{booktabs}
\usepackage[figuresright]{rotating}
\usepackage{makecell}

\makeatletter

\newcommand{\Rmnum}[1]{\expandafter\@slowromancap\romannumeral #1@}
\makeatother

\shorttitle{M Subdwarf Research \Rmnum{3}.}
\shortauthors{Zhang et al.}

\begin{document}
\begin{CJK*}{UTF8}{gbsn}

\title{M Subdwarf Research \Rmnum{3}.  Spectroscopic Diagnostics for Breaking Parameter Degeneracy}

\correspondingauthor{Shuo Zhang \& Hua-Wei Zhang}
\email{szhang0808@pku.edu.cn; zhanghw@pku.edu.cn}

\author[0000-0003-1454-1636]{Shuo Zhang (张硕)}
\affiliation{Department of Astronomy, School of Physics, 
Peking University, Beijing 100871, P. R. China. }
\affiliation{Kavli institute of Astronomy and Astrophysics, 
Peking University, Beijing 100871, P. R. China. }

\author[0000-0002-7727-1699]{Hua-Wei Zhang (张华伟)}
\affiliation{Department of Astronomy, School of Physics, 
Peking University, Beijing 100871, P. R. China. }
\affiliation{Kavli institute of Astronomy and Astrophysics, 
Peking University, Beijing 100871, P. R. China. }

\author{Georges Comte}
\affiliation{Aix-Marseille Univ, CNRS, CNES, LAM, 
Laboratoire d'Astrophysique de Marseille, Marseille, France. }

\author[0000-0002-8546-9128]{Derek Homeier}
\affiliation{Zentrum für Astronomie der Universität Heidelberg, 
Landessternwarte,  Königstuhl 12,  D-69117 Heidelberg,  Germany}

\author[0000-0001-6767-2395]{Rui Wang (王瑞)}
\affiliation{CAS Key Laboratory of Optical Astronomy, 
National Astronomical Observatories,  Beijing 100101,  China}

\author[0000-0001-5541-6087]{Neda Hejazi}
\affiliation{Department of Physics and Astronomy, the University of Kansas, Lawrence, KS 66045, USA}
\affiliation{Department of Physics and Astronomy, Georgia State University, Atlanta, GA 30303, USA}

\author[0000-0001-7607-2666]{Yin-Bi Li (李荫碧)}
\affiliation{CAS Key Laboratory of Optical Astronomy, National Astronomical Observatories, Beijing 100101, China}

\author[0000-0001-7865-2648]{A-Li Luo (罗阿理)}
\affiliation{CAS Key Laboratory of Optical Astronomy, 
National Astronomical Observatories, Beijing 100101, China}
\affiliation{University of Chinese Academy of Sciences, 
Beijing 100049, China}

\begin{abstract}
To understand the parameter degeneracy of M subdwarf spectra at low resolution, we assemble a large number of spectral features in the wavelength range 0.6-2.5 $\mu$m with bandstrength quantified by narrowband indices.  Based on the index trends of BT-Settl model sequences, we illustrate how the main atmospheric parameters ($T_{\rm eff}$,  log $g$,  [M/H] and [$\alpha$/Fe]) affect each spectral feature differently.  Furthermore, we propose a four-step process to determine the four parameters sequentially, which extends the basic idea proposed by Jao et al.  Each step contains several spectral features that break the degeneracy effect when determining a specific stellar parameter.  Finally, the feasibility of each spectroscopic diagnostic with different spectral quality is investigated.  The result is resolution-independent down to R$\sim$200.
\end{abstract}


\section{Introduction} \label{sec:sec1}

Low-mass stars ($M<$ 0.8$M\odot$) are the main stellar component of the Milky Way.  They account for 70\% of the number of stars and occupy 40\% of the total stellar mass of the Galaxy \citep{1997AJ....114.1992R,2003PASP..115..763C,2006AJ....132.2360H,2010AJ....139.2679B,2015AJ....149....5W,2021A&A...650A.201R}.  The M-type low-mass stars at the end of the main sequence in the H-R diagram contain a variety of exciting stellar components, including the most dominant stellar members of Galactic disk - M dwarfs, the rare Population \Rmnum{2} stars - M subdwarfs, and some of the substellar objects - degenerate brown dwarfs.  M dwarfs have contributed significantly to studies of the initial mass function (e.g. \citealt{2012ApJ...747...69C,2016ApJ...821...39M}) as well as the mass-to-light ratio of nearby galaxies \citep{2015ApJ...801...18M, 2015MNRAS.452L..21S}, and they are also popular candidate hosts with earth-sized planet orbiting within the habitable zone (e.g., \citealt{2013ApJ...767...95D}). Their metal-poor counterparts, subdwarfs that associated kinematically with the thick disk and halo (e.g., \citealt{2013AJ....145...40B}), are of importance as the probes of the old galactic populations. Besides,  young brown dwarfs with mass just below the hydrogen-burning minimum mass are sometimes classified to spectral types M7-M9,  mixed with the ultracool dwarfs/subdwarfs with mass $\leq$ 0.1$M\odot$ \citep{2017ApJS..231...15D,2019MNRAS.489.1423Z}.

As the members of old Galactic populations: old disk, thick disk, halo, bulge (see e.g. \citealt{1976ApJ...210..402M,2007ApJ...669.1235L,2013AJ....145...40B,2019AJ....157...63K}),  M subdwarfs are rare in the solar neighborhood and are on average much older than the M dwarfs with solar-metallicity.  The surface chemical compositions of these unevolved main-sequence stars are not changed by various enrichment processes and remain the chemical footprint of the gas from which they formed (e.g. \citealt{2022ApJ...927..122H}).  This makes them fossil record and golden tracers of the earliest phases of the assembly of the Milky Way.

Measuring basic atmospheric parameters is challenging for M-type low-mass stars, because the spectra of these stars with cool atmosphere (2500-4000 K) are dominated by molecular absorption bands,  hiding or blending with the atomic lines, which leaves no windows onto the continuum \citep{1990PhDT.......108A,2014A&A...564A..90R,2016A&A...587A..19P}.  For M dwarfs, stellar properties have been studied relative thoroughly and generally determined separately.  For example, effective temperature ($T_{\rm eff}$) of M dwarfs can be derived from empirical $T_{\rm eff}$-color relationship (e.g., \citealt{2015ApJ...804...64M, 2019ApJ...871...63M}) or $T_{\rm eff}$-index relationship \citep{2012ApJ...748...93R}, while metallicity ([Fe/H]) can be estimated via relations of NIR K-band magnitudes alongside optical photometry \citep{2005A&A...442..635B,2012A&A...538A..25N} and calibrated by binaries containing an FGK-type companion (e.g., \citealt{2005A&A...442..635B,2013AJ....145...52M,2021MNRAS.504.5788R}).  Interferometric diameters and dynamical masses have been utilized to calibrate mass and radius relations (e.g.  \citealt{2000A&A...364..217D,2016AJ....152..141B}).  In some works, comparing synthetic spectra over broad spectral ranges \citep{2013A&A...556A..15R,2014ApJ...791...54G,2021RAA....21..202D} or a series of selected feature bands \citep{2016A&A...587A..19P,2018A&A...615A...6P} to determine several parameters at once is opted for.

However, for metal-poor subdwarfs, the situation is more complicated, because metallicity severely affects the energy distribution in low-mass stars \citep{2000A&A...364..217D}.  Due to the scarcity in the solar neighborhood, it is difficult to obtain enough binaries containing metal-poor FGK star companion that can cover the entire metal abundance range. The low intrinsic brightness of subdwarfs further makes it difficult to acquire adequate high-resolution spectra covering the extended parameter grid, resulting in a poor constraint to the theoretical models.  The near 1-to-1 mass-radius relation for the low-mass dwarfs \citep{2016AJ....152..141B} is no longer effective because metal-deficiency modifies the equilibrium configuration of atmosphere of subdwarfs, leading to smaller radii at the same temperature \citep{2019AJ....157...63K}.  In addition, it is hard to obtain high-quality spectra in the optical because their spectral energy distributions peak at 0.8-1 $\mu$m, while the infrared spectra are seriously contaminated by telluric absorption. To date, comparing observed spectra with grids of synthetic spectra, using $\chi^2$ minimization in multi-dimensional parameter space is still the preferred method \citep{2013A&A...556A..15R,2014A&A...564A..90R,2016A&A...596A..33R,2019A&A...628A..61L,2021ApJ...908..131Z,2020AJ....159...30H,2022ApJ...927..122H}.

When estimating parameters through a synthetic fitting process, parameter degeneracy usually occurs as a problem, because the strength of molecular features is a function of both $T_{\rm eff}$ and metallicity \citep{2021MNRAS.504.5788R}.  For the metal-poor objects, the enhancement of $\alpha$-element (Ne, Mg, Si, S, Ar, Ca, Ti) also impacts the spectral shape over a wide range of wavelength regions \citep{2022ApJ...927..122H}.  According to \cite{2022ApJ...927..122H} who conducted a study on pairwise degeneracy of $T_{\rm eff}$, log $g$, [M/H], and [$\alpha$/Fe], the effects on the spectrum from increasing metal abundance within a certain range can be counteracted by the effect of an increase in $T_{\rm eff}$, an increase in surface gravity or a decrease in [$\alpha$/Fe].  Therefore,  uncertainty is introduced in the fitting process, due to a series of synthetic spectra with similar spectral morphology but different parameter combinations.

In this work, we aim to conduct an extended exploration of spectral degeneracy for M subdwarfs  in the optical and near-infrared when more than two parameters are involved.  The paper is organized as follows.  Section \ref{sec:sec2} explores parameter degeneracy at various feature bands via spectral indices of model sequences.  To break degeneracy,  a four-step process for sequentially estimating $T_{\rm eff}$, [M/H],  [$\alpha$/Fe] and log $g$ is proposed.  Section \ref{sec:sec3} discusses the effect from different spectral quality and resolution.  Finally,  we summarize our study in section \ref{sec:sec4}.

\section{Spectroscopic Diagnostics Determination}\label{sec:sec2}

In the optical and near-infrared spectra of low-mass stars, the most significant opacity sources are metal oxide species such as TiO and CO,  hydrides such as SiH, CaH, FeH, CrH,  hydroxides such as CaOH, and water vapor \citep{2013A&A...556A..15R, 2016A&A...596A..33R}.  The molecular absorption features are consist of thousands of individual lines,  affecting both the detailed structure of the spectrum and the global structure of the atmosphere \citep{1998ApJ...498..851V},  blending to overlapped absorption bands in the low-resolution spectra. 

Narrowband indices were usually designed to estimate the strengths of individual spectral features,  which measure the flux ratios between the feature bands and sidebands (``pseudo-continuum'').  Utilizing narrowband indices with a detailed understanding of corresponding spectral features, one can design effective schemes in estimating parameters.  For example, \cite{1976A&A....48..443M} predicted that TiO absorption decreases in strength with decreasing [M/H] but the hydride bands are largely unaffected, this qualitative result was quantified by \cite{1997AJ....113..806G}, who used several indices (CaH1, CaH2, CaH3, TiO5) to measure the strengths of CaH and TiO bands and developed the first subdwarf classification system.  It is worth noting that then a parameter $\zeta_{\rm TiO/CaH}$ was introduced by \cite{2007ApJ...669.1235L} to quantify the weakening of the TiO band strength.  Its relationships with [Fe/H] and [M/H] + [$\alpha$/Fe] were determined by \cite{2009PASP..121..117W} and \cite{2020AJ....159...30H}, respectively.

The precondition of breaking degeneracy is a detailed understanding of the complicated dependence of more spectral features on atmospheric parameters.  In this section, we propose a solution to parameter measurement, following and extending the basic idea of \cite{2008AJ....136..840J} who proposed a 3-step method to break the degeneracy.  The parameter degeneracy effect is further and deeply explored based on multiple spectral indices.  Note that the analysis of the index trend is based on the synthetic spectra and hence influenced by the incompleteness of models.  Nevertheless,  these trends can still play a theoretical guiding role.

\subsection{PHOENIX BT-Settl Model Grid}

In the present  study,  we have used the latest BT-Settl CIFIST stellar atmosphere models \citep{2012RSPTA.370.2765A,2013MSAIS..24..128A,2014ASInC..11...33A,2015A&A...577A..42B} that also used in \cite{2020AJ....159...30H,2022ApJ...927..122H}.  Compared with the classical grids available from the CIFIST project\footnote{https://phoenix.ens-lyon.fr/Grids/BT-Settl/CIFIST2011/}, this newly calculated model grid\footnote{These models have not yet been made publicly available by the team.} also varies over a range of alpha-element enhancements, [$\rm \alpha$/Fe], as a subgrid. These atmosphere models are computed with the PHOENIX multi-purpose atmosphere code version 15.5 \citep{1997ApJ...483..390H,2001ApJ...556..357A}, including specialized models for the coolest (below 3000 K) stellar and brown dwarf atmospheres using the Settl model of cloud formation as well as the radiation hydrodynamic simulations of M-L-T dwarfs atmospheres \citep{2010A&A...513A..19F, 2012JCoPh.231..919F}. 

Since the release of the BT-Settl model atmospheres,  the pre-calculated synthetic spectra have been widely used in numerous spectroscopic analyses with observations and measure the atmospheric parameters (e.g. \citealt{2013A&A...556A..15R,2014A&A...564A..90R,2016A&A...596A..33R,2018A&A...620A.180R,2018A&A...610A..19R,2013ApJ...779..188M,2015ApJ...804...64M,2017MNRAS.468..261Z,2017MNRAS.464.3040Z,2017ApJ...851...26V,2019MNRAS.489.1423Z,2020AJ....159...30H, 2022ApJ...927..122H,2021ApJ...908..131Z,2021AJ....161..172D}). The results show that BT-Settl models are successful in reproducing the overall optical-NIR spectral profile of M and L subdwarfs, particularly at [Fe/H]$\leq\ -$1.0 dex \citep{2017MNRAS.464.3040Z}, and most of the molecular and atomic features can be well fitted with the observations \citep{2014A&A...564A..90R}. 

In this work, we use the synthetic spectra instead of observed spectra for the subsequent investigation,  because the parameter space of the model grid is uniform and extended to metallicities as low as [M/H] = $-$3.0 dex,  required for our analysis.  The variation of a spectrum solely caused by changing  atmospheric parameters can be explored.  The parameter space is shown in Table \ref{tab:param_grid} and the pre-calculated synthetic spectra have been convolved down to R$\sim$2000 in the following analyses.  At this resolution,  the uncertainty introduced by the imperfection of the models can be referred to the quantitative results from \cite{2020AJ....159...30H,2022ApJ...927..122H},  in which the authors measured that the maximum discrepancies between observation and best-fit synthetic spectra are 5\%-15\% depending on the temperature range and wavebands.

\begin{deluxetable}{ccc}[htb!]
\tablecaption{Parameter space of the model grid used in this work \label{tab:param_grid}}
\tablecolumns{3}
\tablewidth{0pt}
\tablehead{
\colhead{Variable} &
\colhead{Range} &
\colhead{Step size}
}
\startdata
\textit{T}$_{\text{eff}}$& 2500 - 4000 K		& 100 K 			\\              
log \textit{g}  				& 4.5 - 5.5 dex   				&  0.5 dex      	\\
$[\text{M/H}]$  			& $-$3.0 - +0.5 dex 		&  0.5 dex  		\\
$[\alpha/\text{Fe}]$  	& $-$0.2 - 0.2 dex for [M/H] $\geq$ 0.0 & 0.2 dex\\
 & 0.0 - 0.4 dex for [M/H] = $-0.5$ & \\ 
 & 0.2 - 0.6 dex for [M/H] $\leq\ -$1.0 &\\
\enddata
\end{deluxetable}

\subsection{Exploration of Temperature Indicators}

To serve as a qualified temperature indicator,  spectral feature with such characteristics is expected: varying regularly and monotonically with the temperature and almost independent of the effects from any other atmospheric parameter.  In the following, we examine and discuss a class of indices with such properties---pseudo-continuum colors---in a great detail.

\subsubsection{Pseudo-continuum colors}

In general,  the overall optical-to-NIR spectrum is largely depressed by molecular opacity in stars as cool as M subdwarfs,  resulting the true stellar continuum can not be identified.  However, at a few wavelength points,  the molecules are a little more transparent, and one can see deeper in the photosphere, forming a pseudo-continuum \citep{1991ApJS...77..417K,1996ApJ...469..706M}.  Therefore, ``pseudo-continuum colors''  have been defined to estimate the slope of the pseudo-continuum wavelength regions \citep{1993AJ....105.1855H,1996ApJ...469..706M,1999AJ....118.2466M,2002AJ....123.3409H, 2003AJ....125.1598L, 2007AJ....134.2398C, 2014AJ....147...33Y}.

We have collected 16 such ``colors'' from the literature and examined the dependence of each of them on atmospheric parameters.  The pseudo-continuum colors are defined as 
\begin{equation}\label{equa:calculate}
\rm{Color} = \frac{Average\ Flux\ Density\ (Numerator)}{Average\ Flux\ Density\ (Denominator)}
\end{equation}
where Numerator and Denominator are spectral regions within the reference wavelengths listed in Table \ref{tab:color}. The results show that some of these colors are almost insensitive to metallicity and surface gravity, which makes them strong competitors for the temperature calibrators of subdwarfs.

\begin{figure*}[htp!]
\center
\includegraphics[width=120mm]{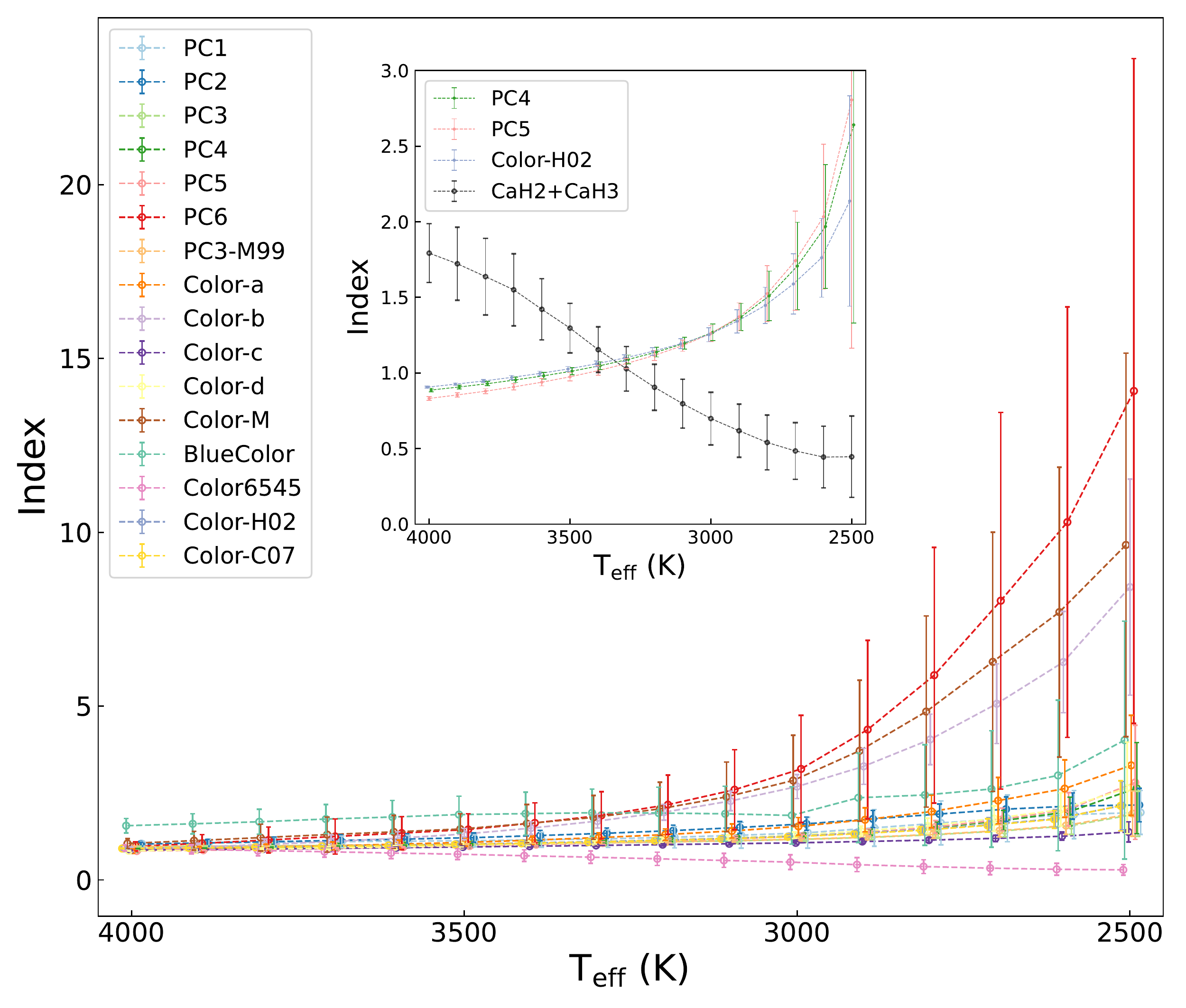}
\caption{Main panel: the value dispersion of 16 psuedo-continuum colors defined in the literature.  The entire model grid from 2500-4000 K with all available gravity, metallicity, and alpha enhancement values are used to calculate.  The synthetic spectra are grouped by temperature, and the mean and standard deviation (shown as the error bar) of each group are calculated for each color.  Inner panel: The three selected pseudo-continuum colors are shown comparing with the composite index CaH2+CaH3.  Their typical dispersion values for different temperatures are listed in Table \ref{tab:dispersion}. \label{fig:PCs}}
\end{figure*}

To measure the variable value range of each color at different temperatures as closely as possible to real conditions,  we select the entire model grid from 2500-4000 K with all available gravity,  metallicity,  and alpha enhancement values listed in Table \ref{tab:param_grid}, and measure the pseudo-continuum colors for each of these synthetic spectra.  We then group the synthetic spectra by temperature, calculate the mean and standard deviation of all colors associated with the spectra in each group, as compared in Figure \ref{fig:PCs}.

From the colors shown in the main panel of Figure \ref{fig:PCs}, we find three colors with a quite small dispersion for each temperature group,  and show them in the inner subfigure of Figure \ref{fig:PCs}. The trend of the commonly used spectral typing indicator,  CaH2+CaH3  \citep{2007ApJ...669.1235L},  is also demonstrated in this subfigure for comparison.  Due to the exclusive dependence of the three pseudo-continuum colors, PC4, PC5, and Color-H02, on the temperature when $T_{\rm eff}\geq$ 3000 K, we choose them to be the temperature indicators.

\subsubsection{Two additional pseudo-continuum colors}\label{subsec:new-colors}

\begin{figure*}[htb!]
\center
\includegraphics[width=120mm]{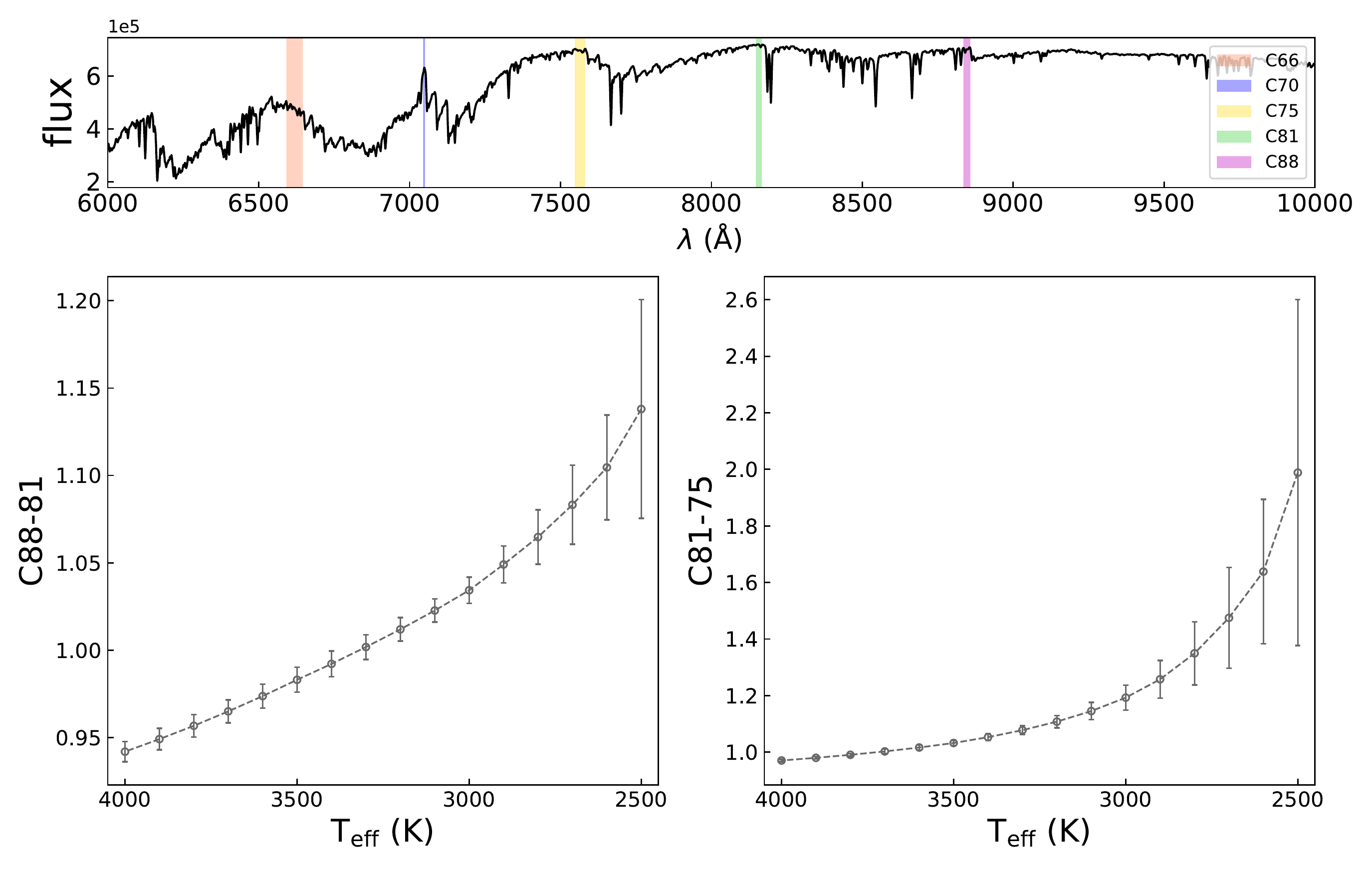}
\caption{Upper panel: the bands listed in Table \ref{tab:mybands} shown on a synthetic spectrum of 3500 K.  Bottom panels: two new pseudo-continuum colors defined in section \ref{subsec:new-colors} that are almost only sensitive to the effective temperature at $T_{\rm eff} \geqslant$ 3000 K.  Table \ref{tab:dispersion} lists their typical dispersion values for different temperatures.\label{fig:mycolors}}
\end{figure*}

Considering that all three selected colors have a reference band beyond 9000 $\rm\AA$ where the observed spectra are dominated by strong telluric absorptions,  we have  further explored alternatives at bluer wavelengths. For this purpose,  we have defined five bands listed in Table \ref{tab:mybands} pertaining to pseudo-continuum points within 6000-9000 $\rm\AA$ which are less depressed by molecular opacity.  Computed from every two bands,  a pseudo-continuum color can measure the slope of the pseudo-continuum within the corresponding wavelength ranges.  The upper panel of Figure \ref{fig:mycolors} shows these bands on a synthetic spectrum, as an example.

\begin{deluxetable}{cccc}[htb!]
\tablecaption{Pseudo-continuum bands defined in this work. \label{tab:mybands}}
\tablecolumns{4}
\tablewidth{0pt}
\tablehead{
\colhead{Band} &
\colhead{Name} &
\colhead{ $\rm \lambda_{begin} (\AA)$} &
\colhead{$\rm \lambda_{end} (\AA)$}
}
\startdata
1& C66 & 6590 & 6645 \\
2& C70 & 7042 & 7049 \\
3& C75 & 7545 & 7580 \\
4& C81 & 8145 & 8165 \\
5& C88 & 8833 & 8855 \\
\enddata
\tablecomments{\footnotesize The colors are named ``CYY-XX'' where YY and XX each represent a reference band and calculated following Equation \ref{equa:calculate}, e.g., C81-66 = $\frac{\text{Average Flux within Band 4}}{\text{Average Flux within Band 1}}$.}
\end{deluxetable}

\begin{deluxetable*}{ccccccccccccccccc}[!]
\center
\tablecaption{Typical dispersion for different effective temperatures \label{tab:dispersion}}
\tablecolumns{17}
\tabletypesize{\tiny}
\tablewidth{7mm}
\tablehead{
\colhead{\normalsize Index} &
\colhead{\scriptsize \makecell{\\4000}} &
\colhead{\scriptsize \makecell{\\3900}} &
\colhead{\scriptsize \makecell{\\3800}} &
\colhead{\scriptsize \makecell{\\3700}} &
\colhead{\scriptsize \makecell{\\3600}} &
\colhead{\scriptsize \makecell{\\3500}} &
\colhead{\scriptsize \makecell{\\3400}} &
\colhead{\scriptsize \makecell{\small{$T_{\rm eff}$ (K)}\\3300}} &
\colhead{\scriptsize \makecell{\\3200}} &
\colhead{\scriptsize \makecell{\\3100}} &
\colhead{\scriptsize \makecell{\\3000}} &
\colhead{\scriptsize \makecell{\\2900}} &
\colhead{\scriptsize \makecell{\\2800}} &
\colhead{\scriptsize \makecell{\\2700}} &
\colhead{\scriptsize \makecell{\\2600}} &
\colhead{\scriptsize \makecell{\\2500}} 
}
\startdata
CaH2+CaH3& 	0.1947	& 0.2412	& 0.2536	& 0.2383	& 0.2019& 	0.1640& 	0.1507	& 0.1470	& 0.1524& 	0.1618& 	0.1739	& 0.1759& 	0.1806	& 0.1878	& 0.2033	& 0.2690\\
PC4& 	0.0120	& 0.0125	& 0.0142& 	0.0181	& 0.0215	& 0.0233	& 0.0265	& 0.0291& 	0.0320	& 0.0394	& 0.0547	& 0.0901	& 0.1641	& 0.2892	& 0.4100	& 1.3107\\
PC5& 	0.0117	& 0.0135& 	0.0162& 	0.0199	& 0.0242& 	0.0282	& 0.0313	& 0.0327	& 0.0334	& 0.0373	& 0.0522	& 0.0924& 	0.1840& 	0.3280& 	0.4790	& 1.6421\\
Color-H02& 	0.0066	& 0.0079	& 0.0099	& 0.0129	& 0.0151	& 0.0168	& 0.0194	& 0.0219	& 0.0255	& 0.0334	& 0.0462	& 0.0767	& 0.1199	& 0.1996	& 0.2608	& 0.6954\\
C81-75	& 0.0047	& 0.0042	& 0.0050	& 0.0082	& 0.0091	& 0.0091	& 0.0121	& 0.0160& 	0.0216& 	0.0306	& 0.0437	& 0.0668	& 0.1116	& 0.1786	& 0.2551	& 0.6115\\
C88-81& 	0.0058	& 0.0062	& 0.0064	& 0.0066	& 0.0068	& 0.0072	& 0.0073	& 0.0071	& 0.0067	& 0.0067	& 0.0075	& 0.0106	& 0.0155	& 0.0226	& 0.0300	& 0.0626\\
\enddata
\tablecomments{\footnotesize The typical dispersion values for the 5 pseudo-continuum colors at different effective temperatures. The classical compound index CaH2+CaH3 is presented for comparison.}
\end{deluxetable*}

After the exploration of all possible colors formed by the above five bands, we find that the pseudo-continuum slopes between the last three bands are mainly dependent on temperature, rather than other parameters.  The dispersion of the pseudo-colors C88-81 and C81-75 are shown in the bottom panels of Figure \ref{fig:mycolors}.  C88-81 has a regular dependence on the temperature, with a small and constant dispersion down to $\sim$3000 K.  Note that the relatively small color range makes it a temperature indicator that would be subject to measurement uncertainties.  On the other hand, C81-75 has a larger dynamic range,  although its increasing dispersion towards lower temperatures may limit its applicability.

\begin{figure*}[htb!]
\center
\includegraphics[width=180mm]{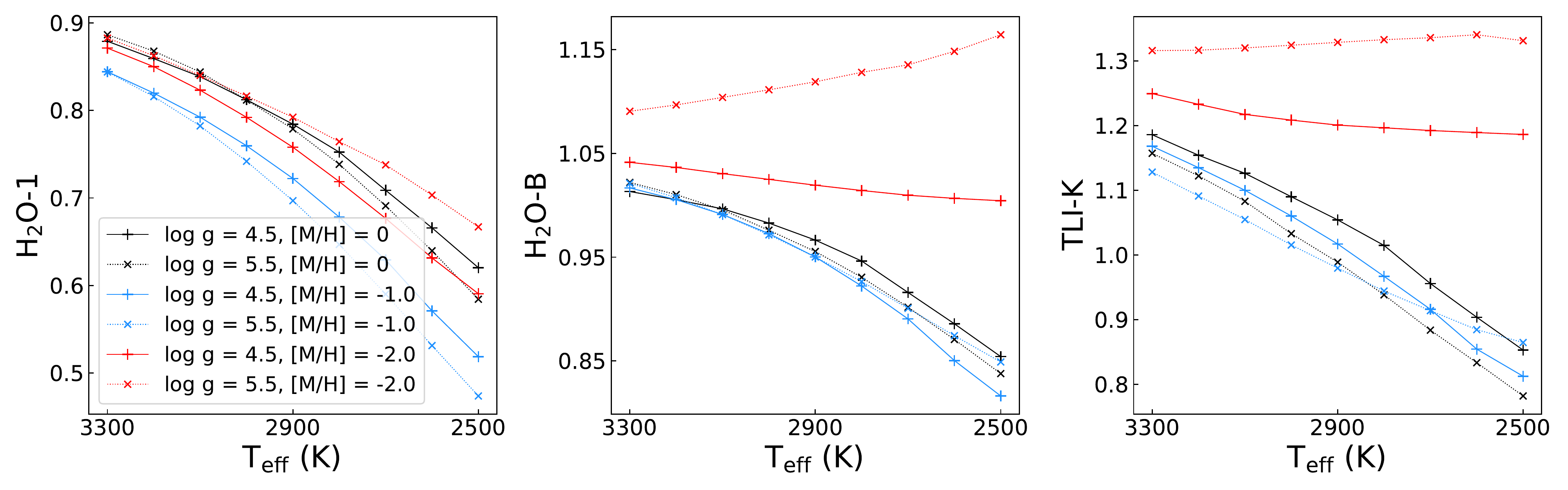}
\caption{Infrared indices as temperature indicators for ultracool subdwarfs.  Models are plotted with the following codes: black refers to solar metallicity with [$\alpha$/Fe] = 0.0,  blue refers to metallicity of $-$1.0 dex with [$\alpha$/Fe] = 0.4 dex,  and red refers to $-$2.0 dex with 0.4 dex;  continuous lines for log $g$ = 4.5 models and dotted lines for log $g$ = 5.5 models.  \label{fig:ultracool}}
\end{figure*}

Caution should be taken when using these colors at low temperatures ($T_{\rm eff}<$ 3000 K), where the color ranges related to the temperature groups rapidly increase.  As temperature decreases below 3000 K, the increasing effect of molecular opacity and cloud formation on stellar atmospheres makes the modeling more complex and difficult \citep{2012RSPTA.370.2765A}.  The optical spectra no longer change sensitively with $T_{\rm eff}$ due to dust formation \citep{2013A&A...556A..15R}.  In addition, the pseudo-continuum colors also begin to be sensitive to other parameters.  If they are still being used as temperature indicators,  much larger uncertainty would thus be introduced.

In the following, we conduct a more extensive investigation of near-infrared spectral features redder than 1 $\mu$m as a  supplement and recommend some of them as temperature tracers for the ultracool subdwarfs.

\subsubsection{Near-infrared temperature indicators }\label{subsubsec:NIR_features}

Spectroscopy at near-infrared wavelengths is involved with a wide range of available atomic and molecular absorption features, especially water vapor absorption bands.  Various studies on late-type M dwarfs and brown dwarfs have defined spectral indices to characterize the molecular bands at different wavelengths, such as H$_2$O, CH$_4$, CO,  and FeH.

We inspect the following 75 indices from the literature characterzing featurebands redder than 1 $\mu$m: 
K1, K2 \citep{1999AJ....117.1010T}, 
Q \citep{2000AJ....119.3019C}, 
water index \citep{2000ApJ...533L..45M},  
H$2$O$^{\text{A}}$, H$2$O$^{\text{B}}$, H$2$O$^{\text{C}}$, H$2$O$^{\text{D}}$ \citep{2001AJ....121.1710R}, 
sHJ, sKJ, sH$_2$O$^{\text{J}}$, sH$_2$O$^{\text{H1}}$,  sH$_2$O$^{\text{H2}}$, sH$_2$O$^{\text{K}}$ \citep{2001ApJ...552L.147T}, 
H$_2$O-A, H$_2$O-B, H$_2$O-C, CH$_4$-A, CH$_4$-B, CH$_4$-C, H/J, K/J, K/H, CO, 2.11/2.07, K shape 
\citep{2002ApJ...564..421B}, 1.0$\mu$m, H$_2$O-1.2, H$_2$O-1.5, CH$_4$-1.6, H$_2$O-2.0, CH$_4$-2.2 \citep{2002ApJ...564..466G},
H$_2$OA, H$_2$OB, H$_2$OC, H$_2$OD, CH$_4$A, CH$_4$B, CO, J-FeH, z-FeH \citep{2003ApJ...596..561M},
H$_2$O-1, H$_2$O-2, FeH \citep{2004ApJ...610.1045S},
z-VO \citep{2005ApJ...623.1115C},
H$_2$O-J, H$_2$O-H, H$_2$O-K, CH$_4$-J, CH$_4$-H, CH$_4$-K, K/J  \citep{2006ApJ...637.1067B}, 
H$_2$O, Na \citep{2007ApJ...657..511A},
WH, WK, QH, QK \citep{2009MNRAS.392..817W},
H-dip \citep{2010ApJ...710.1142B},
H$_2$O-H, H$_2$O-K \citep{2010ApJ...722..971C},
H$_2$O-K2 \citep{2012ApJ...748...93R},
HPI \citep{2012ApJ...744....6S}
FeH$_{\text{z}}$, VO$_{\text{z}}$, FeH$_{\text{J}}$K I$_{\text{J}}$, H-cont \citep{2013ApJ...772...79A},
W$_{\text{0}}$, W$_{\text{D}}$, W$_{\text{1}}$, W$_{\text{2}}$ \citep{2018ApJ...858...41Z},
TLI-J, TLI-K, TLI-g \citep{2022A&A...657A.129A}.

After evaluating the dependence on atmospheric parameters of all indices,  we recommend to use three of them as temperature indicators: H$_2$O-1,  H$_2$O-B,  and TLI-K.  The indices can be calculated as 
\begin{equation}\label{equa:index}
R_{\rm ind} = \frac{F_{\rm W}}{F_{\rm cont}}
\end{equation} 
where pseudo-continuum ($F_{\rm cont}$) region and the feature ($F_{\rm W}$) wavelength limits are listed in Table \ref{tab:features}.

\begin{figure*}[htb!]
\center
\includegraphics[width=180mm]{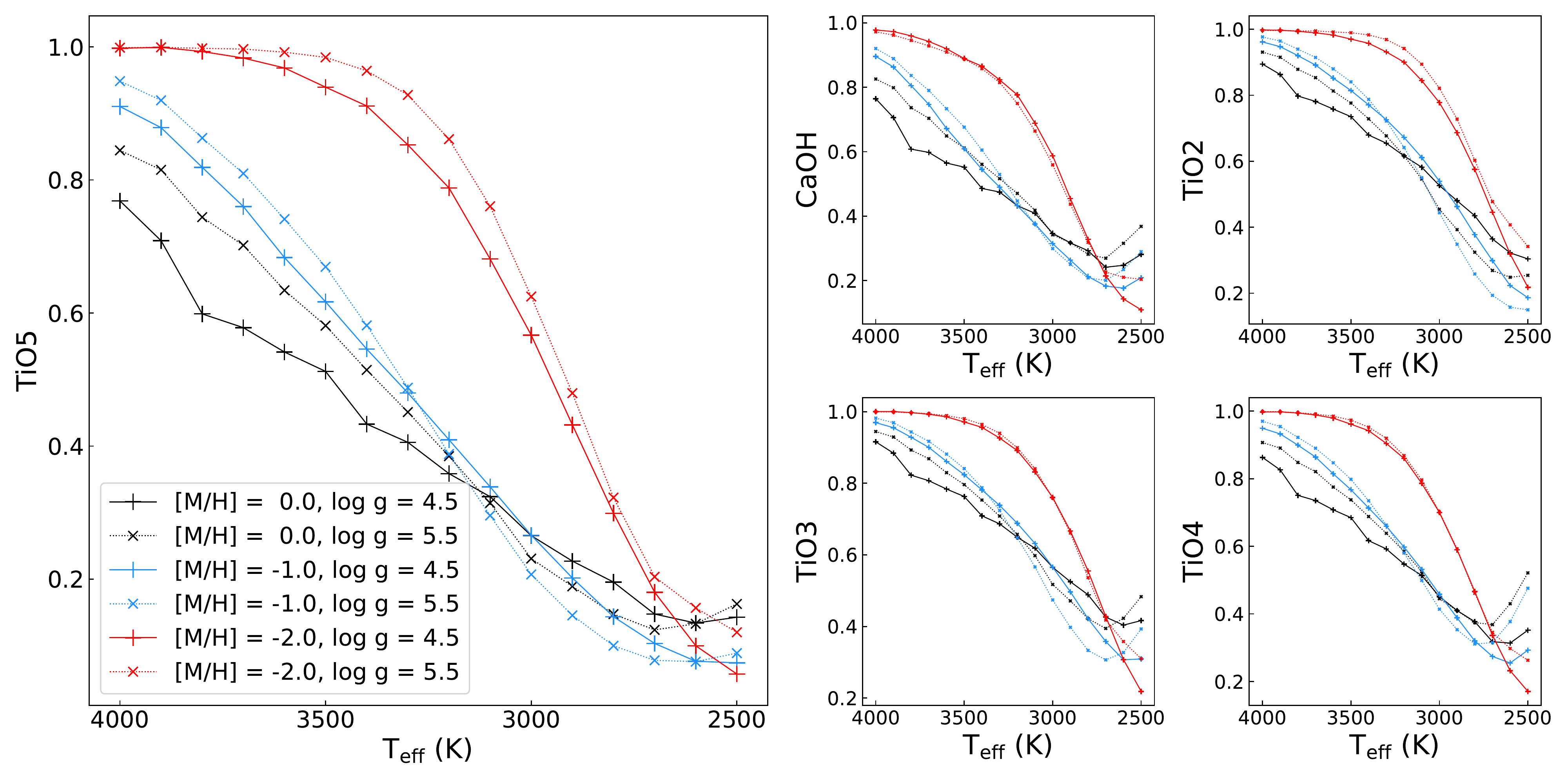}
\caption{The TiO bands represented by the indices of TiO2, TiO3, TiO4 and TiO5 versus the effective temperature. In addition, the CaOH index behaving similarly with the TiO indices is also shown.  Models are plotted with the same codes as in the above figure.  \label{fig:TiO5}}
\end{figure*}

As shown in Figure \ref{fig:ultracool},  the index H$_2$O-1 has a regular temperature dependence at 2500-3500 K.  Note that it also has a modest metal abundance dependence which increases slightly with decreasing temperature. Therefore, two other indices,  i.e., H2O-B and TLI-K,  are supplemented because they present more reliable temperature indicators for solar-abundant to moderately-metal-poor stars (dM/sdM). 

In total,  8 spectral features consisting of 5 pseudo-continuum colors (3 from the literature and 2 defined in this work) and 3 spectral indices (redder than 1 $\mu$m) are provided as qualified temperature indicators.

\subsection{Spectral Features for Estimating [M/H] and [$\alpha$/Fe]}

In the classification criteria of \cite{2008AJ....136..840J}, TiO5 band strength was used to determine metallicity level (see Section 5 in their paper) because the authors claimed that this band strength is highly sensitive to metal abundance variation while almost unaffected by surface gravity.  In the left panel of Figure \ref{fig:TiO5}, we explore the complex dependence of TiO5 on atmospheric parameters based on model sequences with different metal abundances and surface gravity.  Some other indices that have similar behavior, e.g., TiO2, TiO3, TiO4 and CaOH, are also shown in the right panels of Figure \ref{fig:TiO5}.

Two shortcomings limit the reliability of these indices as an indicator of metal abundance, as illustrated in Figure \ref{fig:TiO5}.  First, at high temperatures ($\sim$3300-4000 K), the index values for the solar metallicity to moderate metal-poor models are very sensitive to surface gravity.  As shown in the left panel,  increasing surface gravity and decreasing metal abundance have similar effects on TiO5 index values.  Second,  when the temperature drops below $\sim$3300 K, the index for the models of solar abundance to moderate metal deficiency ([M/H] = $-$1.0 dex) losses its sensitivity to metal abundance, leaving the extremely metal-poor models as the only ones that can be distinguished.  In addition, when the temperature drops below $\sim$2700 K, the index completely loses its ability to determine metal abundance levels due to the disappearance of the TiO molecular bands. 

\begin{figure*}[htb!]
\center
\includegraphics[width=180mm]{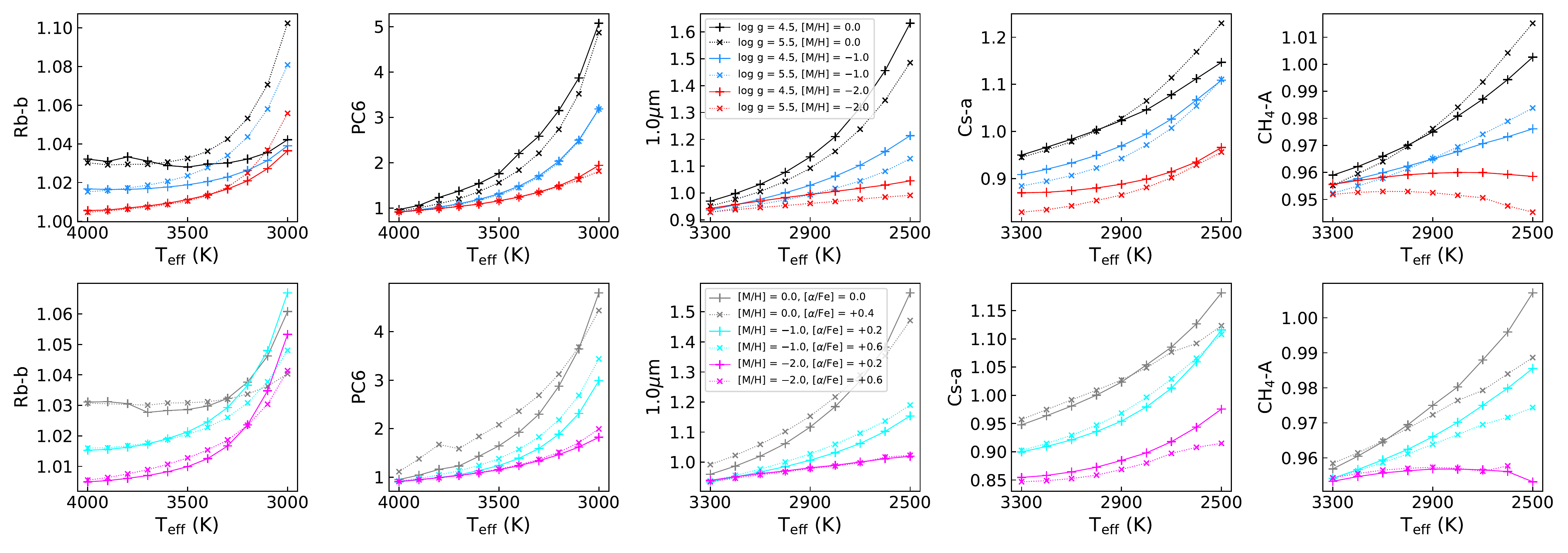}
\caption{Spectral indice with low sensitivity to both surface gravity and alpha enhancement. The upper panels show the model sequences with the same color codes as in the figures above.  The bottom panels show the same indices but of model sequences with log $g$ fixed at 5.0 and varied [$\alpha$/Fe] values. \label{fig:determine-mh}}
\end{figure*}

Besides, TiO is highly sensitive to $\alpha$ abundance, which makes it an indicator of [$\alpha$/H] rather than the overall metallicity [M/H].  Although the values of [$\rm\alpha$/Fe] and [M/H] generally maintain a correlation,  the alpha abundances still differ for every single star and need to be determined individually.  For example, the widely-used model grids such as the classical model grid from the CIFIST project or MARCS model grid \citep{2008A&A...486..951G}  assume [$\rm \alpha$/Fe] as a function of [M/H] based on the rough estimate of [$\rm \alpha$/Fe] for the thin and thick disk.  For this reason, [$\rm \alpha$/Fe] has been considered as a fixed parameter in many studies.  However, large spectroscopic surveys like the Apache Point Observatory Galaxy Evolution Experiment (APOGEE; \citealt{2010IAUS..265..480M,2015ApJ...808..132H}) and the Galactic Archaeology with HERMES (GALAH; \citealt{2021MNRAS.506..150B}) have shown significant scatters around the tight relations assumed in the model grids.  

The $\alpha$ element abundance affects the spectral shape in roughly the same way with [M/H].  An increase in [M/H] may be counteracted by a decrease in [$\alpha$/Fe] and vice versa \citep{2022ApJ...927..122H}.  Therefore, to infer more accurate [M/H], one needs spectral features that are sensitive to neither surface gravity nor $\alpha$ enhancement, but to changes in [M/H] from solar abundance to extreme metal deficiency (provided that the temperature has been already measured).

To search for spectral features with such characteristics, we also explore more spectral features with reference wavebands within 1 $\mu$m defined in the typical studies of M dwarfs/subdwarfs in addition to the NIR indices listed in Section \ref{subsubsec:NIR_features},  including 
CaH 6975, Ti$_{\ \rm \Rmnum{1}}$ 7385, Na$_{\ \rm \Rmnum{1}}$ 8183,8195 \citep{1991ApJS...77..417K}, 
CaOH, CaH1, CaH2, CaH3, TiO1, TiO2, TiO3, TiO4, TiO5 \citep{1995AJ....110.1838R}, 
VO ratio \citep{1995AJ....109..797K}, 
TiO1, TiO2, VO1, VO2, CrH1, CrH2, FeH1,  FeH2, H$_{\rm 2}$O1 \citep{1999AJ....118.2466M}, 
VO-a,  VO-b, Rb-b, Cs-a, CrH-a \citep{1999ApJ...519..802K},
TiO 8440, VO 7434, VO 7912, Na 8190 \citep{2002AJ....123.3409H}, 
TiO6, VO2 \citep{2003AJ....125.1598L}, 
Ca$_{\ \rm \Rmnum{1}}$ 4227, G band, Fe$_{\ \rm \Rmnum{1}}$ 4383, Fe$_{\ \rm \Rmnum{1}}$ 4405, Mg$_{\ \rm \Rmnum{1}}$ 5172, Na D, Mg$_{\ \rm \Rmnum{1}}$ 5172, TiO B, VO 7912, Na$_{\ \rm \Rmnum{1}}$ 8189, Fe$_{\ \rm \Rmnum{1}}$ 8689, Ca$_{\ \rm \Rmnum{2}}$ 8498, Ca$_{\ \rm \Rmnum{2}}$ 8542, Ca$_{\ \rm \Rmnum{2}}$ 8662 \citep{2007AJ....134.2398C}, 
CaH, and TiO 8250 \citep{2014AJ....147...33Y}.

As a result, we select some of these indices for subsequent analysis and list their reference bands in Table \ref{tab:features}.  Five spectral indices exhibiting the expected characteristic within different temperature scopes are shown in Figure \ref{fig:determine-mh}.  In the temperature range of 3000-4000 K, Rb and PC6 show very little sensitivity to both gravity and alpha enrichment at high and low temperature end respectively.  When the temperature drops to  3000 K and even lower,  the indices 1.0$\mu$m, Cs-a and CH$_{\rm 4}$-A show the expected  characteristics. These feature bands can be used to estimate [M/H] after the temperature is determined.

\begin{figure*}[htb!]
\center
\includegraphics[width=150mm]{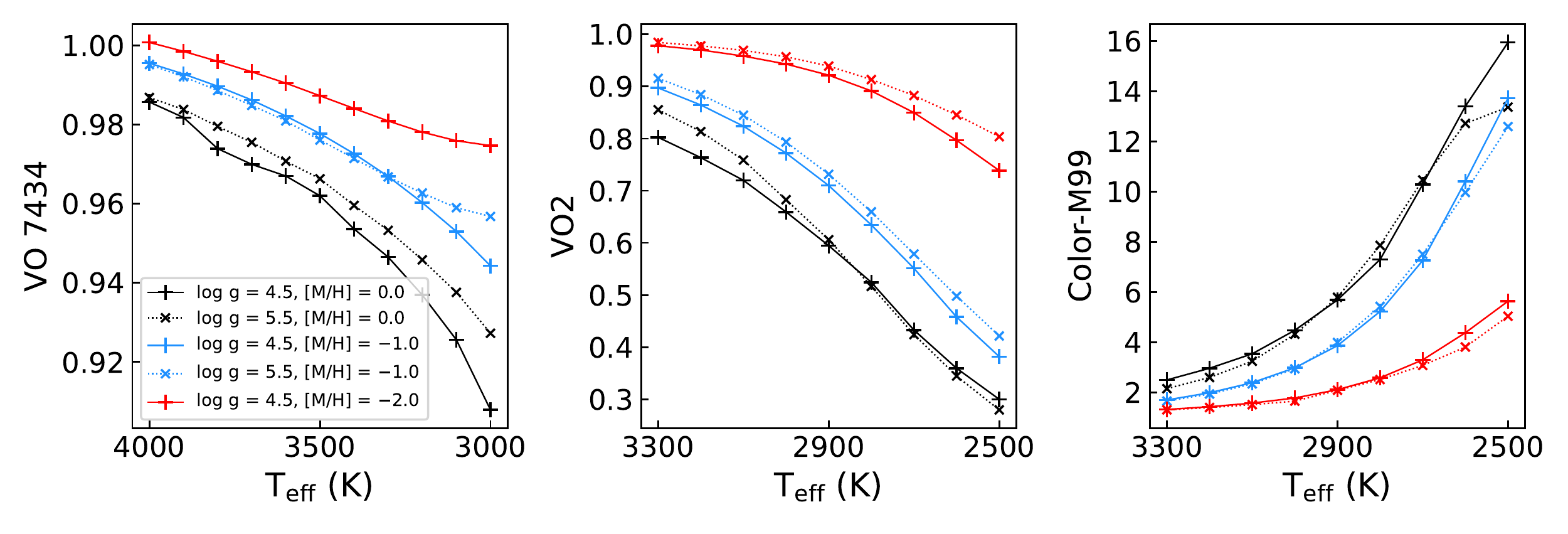}
\caption{Models are plotted with the same color codes as in the above figures. 
\label{fig:determine-alpha}}
\end{figure*}

After [M/H] is determined, spectral features that are sensitive to [$\alpha$/Fe] but still independent of gravity can be used to estimate $\alpha$ abundance. Here we provide three indicators, VO 7434, VO2, and Color-M99, as shown in Figure \ref{fig:determine-alpha}.  VO 7434 exhibits good sensitivity to metal abundance at the high temperature end (3500-4000 K), while VO2 and Color-M99 could well complement its lack of availability due to its increasing sensitivity to gravity at the low temperature end (2500-3500 K).

\subsection{Spectral Features for the Determination of Gravity}

As a final step, spectral features that are highly sensitive to gravity are preferred to determine log $g$.  The classical CaH bandstrengths measured by indices CaH1, CaH2, and CaH3 are dependent on both metallicity and surface gravity.  With the metal abundance determined as a precondition, they could be great indicators of gravity. 

\begin{figure*}[htb!]
\center
\includegraphics[width=180mm]{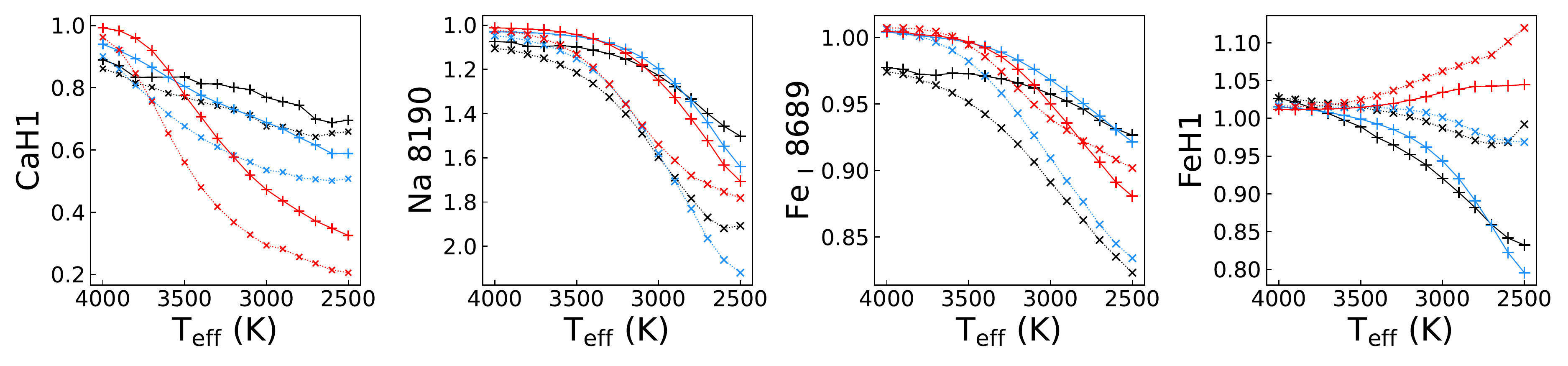}
\caption{Indices that are highly sensitive to gravity and moderately sensitive to metallicity.  Models are plotted with the same color codes as in the above figures. }
\label{fig:determine-logg}
\end{figure*}

In addition to the CaH bands, hydrides such as FeH and the alkali lines such as Na $\rm\Rmnum{1}$ and K $\rm\Rmnum{1}$ doublet are all very sensitive to gravity.  We recommend 4 indices shown in Figure \ref{fig:determine-logg}, i.e.,  CaH1, Na 8190, FeH1  and Fe $_{\Rmnum{1}}$ 8689, because they have a high and regular sensitivity to gravity.  Fe $\rm\Rmnum{1}$ 8689, Na 8190 present a strong gravity dependence for both solar abundance and moderately metal-poor model sequences.  The index FeH1 has a significant dependence to the three parameters, which steadily becomes stronger towards lower temperatures.  Most notably,  the value of this index for the models with [M/H]=$-$2.0 dex increases with decreasing temperature, while the index value for the solar-abundance and moderately metal-poor models decreases.  Besides,  models with high gravity are less sensitive to the temperature except for the extremely metal-poor ones.

\subsection{Parameter estimation steps}\label{subsec:4step}

Accordingly,  we can finally come up with a 4-step solution to the problem due to the degeneracy effect on the parameter determination of M subdwarfs. 
\begin{enumerate}
\item The temperature-sensitive pseudo-continuum colors such as PC4, PC5, Color-H02, C88-81 and C81-75 can be used to estimate the temperature for a subdwarf,  because these indices are hardly affected by the other atmospheric parameters.  For the ultracool objects,  infrared indices H$_2$O-B, and TLI-K are recommended for the dMs and sdMs, while H$_2$O-1 can be used for all the metallicity subclasses as temerature indicators with slightly larger scatters.
\item  Based on the determined temperature,  the spectral index/color Rb-b and PC6 can be used to estimate [M/H] for subdwarfs with temperatures higher than 3000 K because they are not sensitive to either [$\alpha$/Fe] or gravity within given temperature scope.  For subdwarfs with temperatures lower than 3000 K,  the indices 1.0 $\mu$m, Cs-a and CH$_{\rm 4}$-A could be useful.
\item The third free parameter,  alpha enhancement [$\alpha$/Fe], can be estimated from VO 7434, VO2 and Color-M99, as shown in Figure \ref{fig:determine-alpha}, when $T_{\rm eff}$ and [M/H] are known and fixed.
\item Finally, the spectral features that are particularly sensitive to surface gravity will be appropriate to determine log $g$.  The multiple CaH bands (CaH1, CaH2, CaH3) are useful as good indicators,  combining with the features measured by indices Na 8190, Fe $\rm\Rmnum{1}$ 8689 and FeH1.
\end{enumerate}

Particular attention needs to be paid to the fact that each index has its own suitable temperature range and the results of these analyses are all based on low-resolution (R$\sim$2000) synthetic spectra.  

Besides,  considering that the dynamic range of each index is different, and the reference wavelength regions are also affected by observed factors differently, the applicable condition of each index is further discussed in the next section.

\section{Effect From Different Spectral Quality and Resolution}\label{sec:sec3}

The spectral indices were initially defined for low-resolution spectral research because low resolution spectra are readily available and plentiful but lack the detailed structures observed at high resolutions that can be used to accurately infer atmospheric parameters (see e.g. \citealt{2014A&A...564A..90R}).  

In this section, we aim to investigate the effects from reduced spectral qualities which can be quantified by signal-to-noise (S/N) ratios and lower spectral resolutions.

\subsection{Spectral Quality Variations}

All the model sequences in the $T_{\rm eff}$-index diagrams above are based on the analysis results of the synthetic spectra without noise, but in the observations, the noise level of flux directly affects the measurement accuracy of the index. Therefore, we add Gaussian noises corresponding to S/N = 200, 100, 50, 20, 10, and 5 respectively to the synthetic spectra, calculate every index with the errors, and check the lowest S/N ratio that each index can still be used as a spectroscopic diagnostic.

\begin{figure*}[htb!]
\includegraphics[width=180mm]{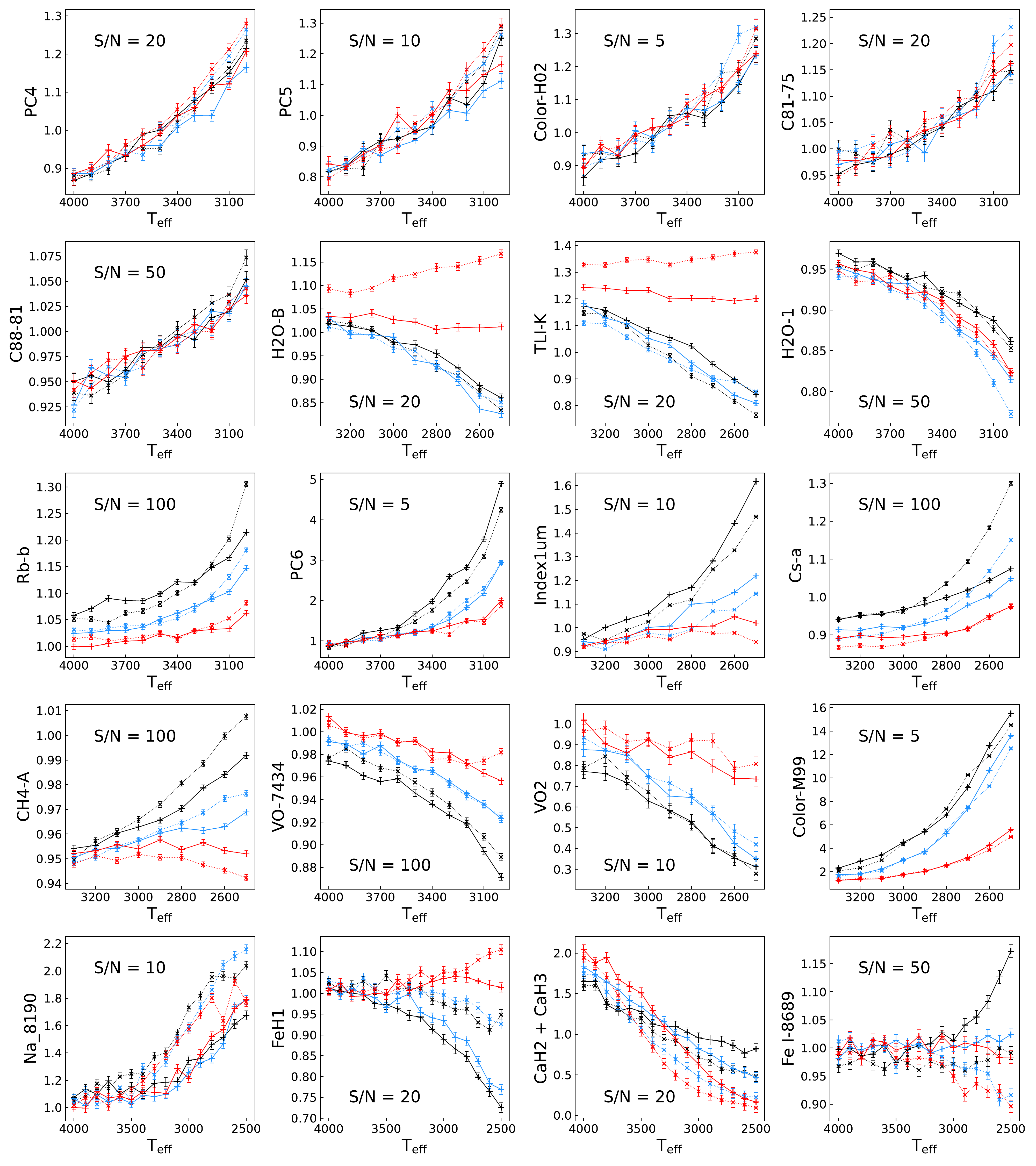}
\caption{Recommended spectral indices derived from spectra with applicable lowest S/N ratios.  The models are plotted with the same color codes as in the above figures.  The error bars of each index representing 1$\sigma$ errors are calculated from synthetic spectra with Gaussian noise added. The corresponding S/N ratio is shown on each sub-diagram. 
\label{fig:multi-sn}}
\end{figure*}

As shown in Figure \ref{fig:multi-sn},  in general, the pseudo-continuum colors are least affected by noise, and some (i.e., Color-H02, PC6, Color-M99) maintain some availability even when the S/N is as low as 5.  On the contrary, a part of the diagnostics for metallicity and alpha abundance, i.e., Rb-b, Cs-a, CH$_{\rm 4}$-A, VO-7434,  depend very much on the quality of the spectrum: only when the S/N ratio exceeds 100, these indices distinguish finely between different abundances.

In addition, the effects that from observation and may introduce uncertainties need to be calibrated carefully in advance, such as telluric line contamination, flux calibration issues, spectral reddening and so on.

\subsection{Lower Spectral Resolutions}

At lower resolutions,  problems could be raised by the information loss and the decreased number of sampling points involved in the index calculation. 

\begin{figure*}[htb!]
\center
\includegraphics[width=150mm]{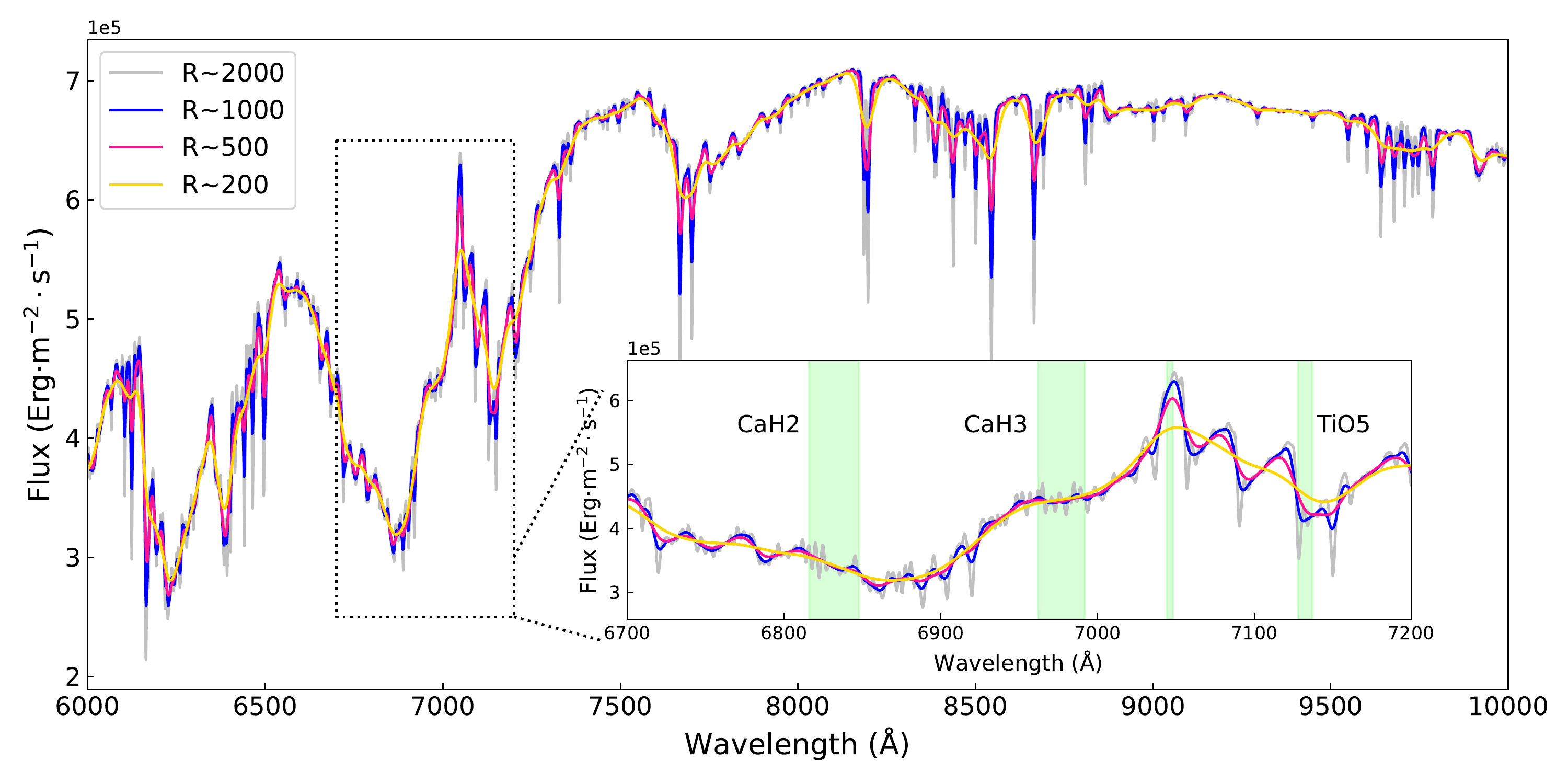}
\caption{A synthetic spectrum ($T_{\rm eff}$ = 3500 K,  log $g$ = 5.0,  [M/H]= $-$1.0 dex,  [$\rm \alpha$/Fe] = 0.4 dex) at different resolutions: R$\sim$2000, 1000, 500, 200, respectively. The spectral region 6700-7200 $\rm \AA$ is enlarged and shown in the inner subfigure. The CaH2, CaH3 and TiO5 feature bands and the reference band that they share are marked with lightgreen shadow.
\label{fig:multi-res}}
\end{figure*}

As the resolution decreases, the overall shape of the SED remains basically unchanged, while the absorption lines become shallower and wider, the line-wings change from sharp to flat, and some more structural characteristics such as the ``jagged'' TiO molecular band near 7000 $\rm\AA$ gradually smooths until almost invisible at R$\sim$200.  Figure \ref{fig:multi-res} shows a comparative example of a synthetic spectrum convolved to different resolutions. 

On the other hand,  the wavelength points within the reference bands of a spectral index decline in the number with the resolution decreasing.  Assuming an observed spectrum is sampled with a proper sampling rate,  e.g., 2.5 times the full width at half maximum (FWHM) of the line spread function (LSF),  some very narrow reference bands may not have any points left to calculate when the resolution drops to R$\sim$500 or lower.  In this situation,  it is necessary to over-sample the observed spectrum before calculating any index.

In order to investigate the changes in the spectral indices caused by different resolutions and explore the applicability of these features, we additionally convolve the synthetic spectra to R $\sim$ 1000, 500,  and 200, respectively. 

The indices that are mostly influenced by the resolution variation are expected to be the absorption atomic lines (such as Na \Rmnum{1} doublet) and the narrow-band molecular features (such as TiO5 and CaH2/3) which has one narrow reference band as shown in the inner subfigure of Figure \ref{fig:multi-res} - it is important to recall that the spectral features explored in this study have been measured by the average or integrated flux within several specific bands.  

Nevertheless, upon closer inspection, we find that the vast majority of index trends of our recommended spectral features are resolution-independent down to R$\sim$200, while they are also affected by noise levels very similarly as at resolution R$\sim$2000.  Since the performance of indices at these low resolutions is basically the same and does not provide significantly more information than Figure \ref{fig:multi-sn}, we do not illustrate further the results here.

\subsection{Discussion on Applicability to Observations}

As a result, the exploration results of this paper can be also applied to very low-resolution spectral data, such as the slitless spectroscopic survey conducted by The Chinese Space Station Telescope (CSST; \citealt{2011SSPMA..41.1441Z, 2018cosp...42E3821Z,2019ApJ...883..203G}) which aims to deliver high-quality spectra covering 2500-10,000 $\rm \AA$ at R$\sim$200 for hundreds of millions of stars and galaxies.  Even so,  it is worth noting that flux calibration with high precision and high spectral quality are the foundation of the accuracy of index measurement.  In many spectroscopic surveys, the flux calibration of cool stellar spectra often suffers larger uncertainty than the hot ones, mainly due to the lack of standard stars and inaccurate extinction correction. 

We want to mention that,  the above-proposed technique to solve the long-standing issue due to the parameter degeneracy is still in its early stage and more careful examinations are needed.  In practice, a preset of initial values for the four parameters is required to start the a fitting process.  We suggest using the approach presented in \cite{2022ApJ...927..122H} (Section 4.2 in their paper) which can allow us to develop an automated pipeline that will be applicable to future spectroscopic surveys.  

In addition,  a well-designed narrow-band photometric survey covering the selected wavelength bands can also be a competitive implementation option.  In recent years, many narrow-band photometric surveys have been successfully carried out, such as the Javalambre Physics of the Accelerating Universe Astrophysical Survey (J-PAS; \citealt{2012SPIE.8450E..3SM,2014arXiv1403.5237B}),  the Javalambre-Photometric Local Universe Survey (J-PLUS; \citealt{2019A&A...622A.176C,2022A&A...659A.181Y,2022arXiv220502595W}), and the Southern Photometric Local Universe Survey (S-PLUS; \citealt{2019MNRAS.489..241M,2022MNRAS.511.4590A}).  Color-H02 (7350-7500 $\rm \AA$ and 8900-9100 $\rm \AA$),  for example,  can be completely covered by filters 38 and 54 of J-PAS which used 56 narrow-band filters to sample the spectral energy distribution in the optical (3800-9200 $\rm \AA$) and achieved 1\% photometric precision. 

\section{Summary and Conclusion}\label{sec:sec4}
In this paper,  we conduct an extended exploration of parameter degeneracy of basic atmospheric parameters ($T_{\rm eff}$, log $g$, [M/H],  and [$\alpha$/Fe]) in the optical to near-infrared spectra of M subdwarfs.  We assemble a large number of pseudo-continumm colors and spectral indices which can quantify the slope of given pseudo-continuum and bandstrength of molecular absorption, respectively.  Based on the index trends of the latest PHOENIX BT-Settl model sequences, we illustrate with figures how the degenerated parameters affect the bandstrength of each spectral feature.

Furthermore, we propose a four-step process (see Section \ref{subsec:4step}) to determine $T_{\rm eff}$, [M/H], [$\alpha$/Fe] and log $g$ sequentially,  which extends the basic idea proposed by Jao et al.  and effectively breaks the degeneracy.  To this end, we suggest several spectral features to be used in each step that determines a specific parameter.  Note that although the suggested features are characterized by indices for our quantified investigation,  the practical way to use these features may include, but not limited to, selecting corresponding wavelength regions for spectral fitting, calculating index values and obtain empirical relationships, and adopting proper corresponding narrow-band photometry instead of spectral data. 

The effect of different spectral quality and resolution are also explored, drawing a conclusion that pseudo-continuum colors are least affected by noise among the recommended spectral features, while some diagnostics for metallicity and alpha abundance depend very much on the quality of the spectrum.  Most of index trends of spectral features that we recommended are resolution-independent down to R$\sim$200,  but they are also affected by noise levels very similarly as at high resolution. Finally, we discuss the possibility of using narrow-band photometry as an alternative option for spectral data.

\begin{acknowledgments}
We thank the anonymous referee for helpful comments and suggestions which significantly improved our manuscript.  Z.S.  thanks Dr.  Fu X.T.  for the thoughtful discussion.  This work was funded by the National Key R$\&$D Program of China (No. 2019YFA0405500), the National Natural Science Foundation  of China (NSFC Grant No. 11973001, 12090040, 12090044,12103068), and the Science Research Grants from the China Manned Space Project with No. CMS-CSST-2021-B05, CMS-CSST-2021-A08.
\end{acknowledgments}

\appendix
\section{Pseudo-continuum colors and spectral indices with their reference bands}

\begin{deluxetable}{ccccc}[htb!]
\tablecaption{The pseudo-continuum colors. \label{tab:color}}
\tablecolumns{5}
\tablewidth{0pt}
\setcounter{table}{0}
\renewcommand{\thetable}{A\arabic{table}}
\tablehead{
\colhead{Pseudo-continuum Color} &
\colhead{Name in this paper} &
\colhead{ $\rm Numerator (\AA)$} &
\colhead{$\rm Denominator (\AA)$}&
\colhead{Source} 
}
\startdata
BlueColor  &  & 6100-6300 & 4500-4700 & \citet{2007AJ....134.2398C}\\
Color6545 & & 6545-6549 & 7560-7564 & \cite{2014AJ....147...33Y} \\
Color-M & & 8105-8155 & 6510-6560 & \citet{2003AJ....125.1598L} \\
Color-1 & Color-H02 & 8900-9100 & 7350-7500 & \citet{2002AJ....123.3409H} \\
Color-1 & Color-C07& 8900-9100 & 7350-7550 & \citet{2007AJ....134.2398C} \\
Color-a & & 9800-9850 & 7300-7350 & \citet{1999ApJ...519..802K} \\
Color-b & & 9800-9850 & 7000-7050 & \citet{1999ApJ...519..802K} \\
Color-c & & 9800-9850 & 8100-8150 & \citet{1999ApJ...519..802K} \\
Color-d & & 9675-9875 & 7350-7550 & \citet{1999ApJ...519..802K} \\
PC1 & & 7030-7050 & 6525-6550 & \citet{1996ApJ...469..706M} \\
PC2 & & 7540-7580 & 7030-7050 & \citet{1996ApJ...469..706M} \\
PC3 & & 8235-8265 & 7540-7580 & \citet{1996ApJ...469..706M} \\
PC3 & PC3-M99& 8230-8270 & 7540-7580 & \citet{1999AJ....118.2466M} \\
PC4 & & 9190-9225 & 7540-7580 & \citet{1996ApJ...469..706M} \\
PC5 & & 9800-9880 & 7540-7580 & \citet{1996ApJ...469..706M} \\
PC6 & & 9090-9130 & 6500-6540 & \citet{1999AJ....118.2466M} \\
\enddata
\tablecomments{\footnotesize{The pseudo-continuum colors assembled from the literature.  Some of the colors are renamed to avoid the problem of duplicate names.}}
\end{deluxetable}

\begin{deluxetable}{cccccc}[htb!]
\tablecaption{The spectral features defined in the literature. \label{tab:features}}
\tablecolumns{6}
\tabletypesize{\scriptsize}
\tablewidth{10mm}
\setcounter{table}{1}
\renewcommand{\thetable}{A\arabic{table}}
\tablehead{
\colhead{Spectral Index} &
\colhead{Name in this paper}&
\colhead{Numerator($\rm\AA$)} &
\colhead{Denominator($\rm\AA$)}&
\colhead{Method}&
\colhead{Source}
}
\startdata
CaOH & & 6230-6240 &  6345-6354 &  average & \citet{1995AJ....110.1838R} \\
CaH1 & & 6380-6390 &  6345-6355, 6410-6420 &average & \citet{1995AJ....110.1838R}\\
CaH2 &  & 6814-6846 & 7042-7046 & average  & \citet{1995AJ....110.1838R}\\
CaH3 & & 6960-6990 & 7042-7046 & average  & \citet{1995AJ....110.1838R}\\
TiO2 & & 7058-7061 & 7043-7046 &  average & \citet{1995AJ....110.1838R} \\
TiO3 & & 7092-7097 & 7079-7084 & average  & \citet{1995AJ....110.1838R} \\
TiO4 & & 7130-7135 &  7115-7120 & average & \citet{1995AJ....110.1838R} \\
TiO5 & & 7126-7135 & 7042-7046 & average & \citet{1995AJ....110.1838R}  \\
VO 7434 & & 7430-7470 &  7550-7570 & average  & \citet{2002AJ....123.3409H}  \\
VO2 & & 7920-7960 &  8130-8150 & average & \citet{2003AJ....125.1598L} \\
Rb-b & & 7922.6-7932.6, 7962.6-7972.6 & 7942.6-7952.6 & average & \citet{1999ApJ...519..802K} \\
Na 8190  & & 8140-8165 & 8173-8210 & average & \citet{2002AJ....123.3409H}\\ 
Fe$_{\ \rm \Rmnum{1}}$ 8689 & & 8684-8694 &8664-8674 &  average & \citet{2007AJ....134.2398C} \\
Cs-a & & 8496.1-8506.1, 8536.1-8546.1 & 8516.1-8526.1 & average & \citet{1999ApJ...519..802K} \\
FeH1 & & 8560-8660 &  8685-8725 & average  & \citet{1999AJ....118.2466M} \\
1.0$\mu$m & & 10,400-10,500  & 8750-8850  & integrated& \citet{2002ApJ...564..466G}\\
CH$_{\rm 4}$-A & & 12,950-13,250  & 12,500~12,800 & average & \citet{2002ApJ...564..421B}\\
H$_2$O-1  & & 13,350-13,450 & 12,950-13,050 & average & \citet{2004ApJ...610.1045S}\\
H$_2$O-B & &15,050-15,250 & 15,750-15,950 & average & \citet{2002ApJ...564..421B} \\
TLI-K && 19,700-19,900 & 22,200-22,400 & average & Almendros-Abad et al. (2022)
\enddata
\tablecomments{\footnotesize Spectral indices assembled from the literature and recommended in the 4-step process.  For each index, one can use the corresponding ``Method'' to calculate the flux over the ``Numerator'' and ``Denominator'' wavelength ranges respectively and derive the index value according to Equation \ref{equa:index}.  In the case of CaH1, Rb-b and Cs-a, two bands in Numerator/Denominator column are both used as the feature/pseudo-continuum.}.
\end{deluxetable}
\clearpage
\bibliography{manuscript}{}
\bibliographystyle{aasjournal}

\end{CJK*}
\end{document}